\documentclass[prl,twocolumn,showpacs]{revtex4}

\usepackage{amsfonts}

\def\bmath#1{\mbox{\boldmath$#1$}}

\def\RR{{\rm\kern.24em \vrule width.04em height1.48ex depth-.07ex
\kern-.30em R}}

\begin{document}

\title{Bell Inequalities with Auxiliary Communication}
\author{D. Bacon}
\email{dabacon@cs.caltech.edu}
\author{B. F. Toner}
\email{toner@theory.caltech.edu}
\affiliation{Institute for Quantum Information, California Institute of
Technology, Pasadena, CA 91125}
\affiliation{Department of Physics, California
Institute of Technology, Pasadena, CA 91125}

\begin{abstract}
  What is the communication cost of simulating the correlations
  produced by quantum theory?  We generalize Bell inequalities to the
  setting of local realistic theories augmented by a fixed amount of
  classical communication.  Suppose two parties choose one of $M$
  two-outcome measurements and exchange one bit of information.  We
  present the complete set of inequalities for $M=2$, and the complete
  set of inequalities for the joint correlation observable for $M=3$.
  We find that correlations produced by quantum theory satisfy both of
  these sets of inequalities, irrespective of the particular quantum
  state or the specific measurements.  One bit of communication is
  therefore sufficient to simulate quantum correlations in both of
  these scenarios.
\end{abstract}


\maketitle

What are the differences between a quantum information processing
device and its classical counterpart?  The discovery that quantum
computers can outperform classical computers~\cite{ShorGrover} has
made answering this question a central goal for the field of quantum
information.  Nearly forty years ago, Bell~\cite{Bell:64a} pointed out
that the correlations resulting from quantum theory cannot be
reproduced by any classical local realistic theory.  It follows that
quantum correlations on spacelike separated systems cannot be
reproduced classically.  If, however, the systems are timelike
separated, then classical simulation is possible, albeit at the
expense of some communication, but {\em how much} is required?  In
particular, suppose a number of spatially separate parties share an
entangled quantum state, and each makes a local measurement on their
component.  Then quantum correlations are manifest in the {\em joint
  probability distribution} of the parties' outcomes, dependent on
each party's choice of measurement.  If this probability distribution
cannot be reproduced by a classical local realistic theory, then it
violates some generalized Bell inequality~\cite{nosignal}.  This means
some communication between the parties is required to reproduce the
probability distribution, but Bell inequality violation does nothing
to {\em quantify} how much. More generally, entanglement is a {\em
  resource} for performing information processing tasks, and an
important goal of quantum information theory is to demarcate the
difference between it and classical resources, such as shared
randomness and classical communication channels.  What classical
resources are required to reproduce the joint probability
distributions arising from local measurement on shared quantum states?

We address the above question in this Letter.  Within the setting of
local realistic theories augmented by a fixed amount of {\em two-way}
classical communication~\cite{Brassard:99a}, we introduce the notion
of {\em Bell inequalities with auxiliary communication}.  These
inequalities provide conditions on the joint probability distribution,
which must be satisfied if such correlations can be simulated with
shared randomness and a fixed amount of communication.  Of particular
significance are complete sets of such inequalities, which provide
necessary and sufficient conditions.  In the scenario where two
parties choose one of $M$ two-outcome measurements and exchange one
bit of information, we present the complete set of inequalities for
$M=2$, and the complete set of inequalities for the joint correlation
observable for $M=3$.  We find that quantum correlations satisfy all
of these inequalities, irrespective of the particular quantum state or
the specific measurements, and can therefore be explained in these
settings by augmenting a local realistic theory with a single bit of
communication.  This is particularly remarkable for the $M=3$ case,
where one would naively expect a trit of auxiliary communication is
required to simulate quantum correlations.

{\em The model.}--- We restrict attention to scenarios with two
parties, $A$ and $B$.  In a {\em measurement scenario} for this
bipartite case, each party selects one of $M$ different measurements
and then---possibly after some delay, during which we might allow the
parties to communicate---outputs one of $K$ different outcomes.  (Note
that $A$ and $B$ may choose measurements from different $M$-element
sets.)  Such a measurement scenario results in a set of probabilities
$0 \leq p_{a,b|i,j} \leq 1$, where $p_{a,b|i,j}$ is the probability
that $A$ outputs $a$ and $B$ outputs $b$, given that $A$ selects
measurement $i$ and $B$ selects measurement $j$.  Discounting null
outcomes (which can be incorporated as a separate outcome if desired),
it follows that $\sum_{a=0}^{K-1}\sum_{b=0 }^{K-1} p_{a,b|i,j}=1$,
where $0 \leq i,j \leq M - 1$.  A valid measurement scenario is any
set of probabilities which satisfy these normalization constraints.

Given a particular measurement scenario, we investigate all {\em
  protocols} which the two parties might perform to produce the
correct probabilities.  A protocol consists of three stages: (i) {\em
  preparation} via the distribution of shared randomness, (ii) {\em
  communication} via the exchange of messages between the parties, and
(iii) {\em output} of outcomes by each party as determined by
information accessible to each party.  $A$ and $B$ select their
measurements after step (i) but before step (ii).  If a protocol
produces identical probabilities to the measurement scenario, then we
say that the protocol has {\em simulated} the scenario.

Two informational resources are of interest: the quantity of shared
randomness and the amount of communication between the parties.  We
focus on the amount of communication and define the {\em cost} of a
protocol to be the maximum amount of communication required (as
opposed to the average amount of communication, see~\cite{average}).
In the preparation phase of a protocol, we allow $A$ and $B$ to share
an infinite amount of classical information and, in particular,
continuous variables.  In the parlance of foundational studies of
quantum theory, these are known as local hidden variables
(LHVs)~\cite{Bell:93a}.  A protocol with no communication (step (ii)
missing) is usually called a {\em LHV theory}.  In such a theory, each
party's output depends on the shared randomness and on which
measurement the particular party has locally selected, but not on the
measurement choice of the other party.

The protocols we investigate are therefore an extension of LHV
theories, where we allow the parties to communicate after selecting
measurements~\cite{Brassard:99a}.  This allows some
``which-measurement'' information to propagate between the parties.
We emphasize that a protocol of this form simulates the joint
probability distribution resulting from a set of quantum measurements,
\emph{but not} the quantum measurements themselves: it is not possible
to replace local measurements made by two spacelike separated parties
on an entangled quantum state by classical communication.  Even in
this case, however, the amount of two-way communication required to
reproduce the joint probability distribution provides a measure of the
{\em nonlocality} of the correlations.  From an information processing
perspective, this model provides a fair setting for the comparison of
quantum correlations and classical resources required to reproduce
them.

Of particular significance in this respect is the result of Brassard,
Cleve, and Tapp~\cite{Brassard:99a}, who demonstrated that the
correlations produced by two-outcome projective measurements on an EPR
pair can be simulated by a local realistic theory augmented by eight
bits of communication.  Surprisingly, we have recently shown that a
single bit of communication is sufficient~\cite{Toner:02a}.

Little, however, is known for more general states and more general
measurements.  The goal of this paper is to illuminate how such bounds
can be achieved by generalizing Bell inequalities to what we term,
{\em Bell inequalities with auxiliary communication}.

{\em Bell polytopes.}---Bell inequalities~\cite{Bell:64a,Clauser:69a}
describe necessary conditions on the probabilities $p_{a,b|i,j}$,
which must be satisfied if these probabilities are to be produced by a
local realistic theory.  When a set of these conditions is also
sufficient, we say that we have a complete set of Bell inequalities.
The construction of complete sets of Bell inequalities is an exercise
in convex geometry~\cite{convex}.  In this section, we briefly sketch
the construction for Bell inequalities without auxiliary
communication.

Consider a deterministic protocol, i.e., one in which no randomness,
shared or otherwise, is used.  (This corresponds to a protocol
consisting only of step (iii) above, with the additional requirement
that this step is deterministic.)  Each party's output can only depend
on their local which-measurement information, so that all such
protocols can be completely characterized by two functions
$\alpha,\beta:{\mathbb Z}_M \to {\mathbb Z}_K$, which describe the
outcomes of the two parties' measurements: if $A$ selects measurement
$i$, she outputs $\alpha(i)$ and if $B$ selects measurement $j$, he
outputs $\beta(j)$.  The probabilities for the scenario are then
$p_{a,b|i,j}=\delta^a_{\alpha(i)} \delta^b_{\beta(j)}$.

By listing the components, we may view the probabilities $p_{a,b|i,j}$
as vectors $\vec{p}$ in ${\mathbb R}^D$ with $D=M^2 (K^2-1)$ (recall
the constraint $\sum_{a,b} p_{a,b|i,j}=1$).  To each pair of functions
$\{\alpha,\beta \}$, there corresponds a deterministic protocol, so
the set of all deterministic protocols is a finite collection of such
vectors $\{\vec{d}_\zeta | \zeta = 1,...,K^{2M} \}$.

Now consider the effect of allowing randomness.  Any unshared
randomness can always be replaced by shared randomness on which the
other party does not act~\cite{Kushilevitz:97a}, so we may continue to
assume step (iii) is deterministic.  But then every set of random
variables in step (i) corresponds to a particular deterministic
protocol.  Therefore the set of {\em all possible protocols which use
  randomness and no communication} is described by a convex sum of the
deterministic protocols without communication
\begin{equation}
\vec{p}= \sum_\zeta \lambda_\zeta \vec{d}_\zeta, \quad \sum_\zeta
\lambda_\zeta =1, \quad \lambda_\zeta \geq 0.
\end{equation}
The set of all protocols therefore corresponds to a region
$\Omega_{MK}$ in ${\mathbb R}^D$, which is a polytope because there
are a finite number of extreme vectors
$\vec{d}_\zeta$~\cite{Rockafellar:70a}.  This permits an alternative
description: instead of describing the polytope $\Omega_{MK}$ as the
convex combination of a finite set of extreme points, we can instead
describe it by specifying a complete (finite) set of facet
inequalities.  A facet inequality is a pair $\{\vec{f}, c\}$ which
defines a half-space of ${\mathbb R}^D$ via the inequality $\vec{f}
\cdot \vec{p} \leq c$.  Complete sets of facet inequalities
$\vec{f}_\eta,c_\eta$ are satisfied if and only if $\vec{p}$ is in
$\Omega_{MK}$:
\begin{eqnarray}
\vec{p} \in \Omega_{MK}~{\rm iff}~\vec{f}_\eta \cdot \vec{p} \leq c_\eta,\
\forall \eta.
\end{eqnarray}
Each facet is therefore a Bell inequality and complete sets of facet
inequalities are complete sets of Bell inequalities. Complete sets are
known in the two party case when $M=2, K=2$~\cite{Fine:82a}, $M=3,
K=2$~\cite{Pitowsky:01a}, and also when extra symmetry constraints are
imposed~\cite{Garg:84a}.

{\em Bell inequalities with auxiliary communication.}---We now turn to
the main focus of our Letter: extending the formalism of Bell
inequalities to protocols which permit communication after the parties
have chosen their measurements.  Again consider a deterministic
protocol, but now allow for the communication (possibly two way) of at
most $r$ bits of information between the parties after selection of
measurements.  Such a protocol is completely characterized by two
functions $\alpha,\beta:{\mathbb Z}_M \otimes {\mathbb Z}_M \to
{\mathbb Z}_K$, which describe the outcomes of the two parties'
measurements, but now each party's output can also depend on which
measurement the other party selects: if $A$ selects measurement $i$
and $B$ measurement $j$, $A$ outputs $\alpha(i,j)$ and $B$ outputs
$\beta(i,j)$.  The probabilities for such a deterministic protocol are
then $p_{a,b|i,j}=\delta_{\alpha(i,j)}^{a}\delta_{\beta(i,j)}^{b}$.
While $\alpha$ and $\beta$ can now depend on which measurement the
other party selects, not all functions $\alpha(i,j)$, $\beta(i,j)$ are
necessarily accessible, if the parties exchange at most $r$ bits of
communication.  The set of possible functions $\alpha(i,j)$,
$\beta(i,j)$ for protocols which use at most $r$ bits of communication
is the subject of the field of communication
complexity~\cite{Yao:79a,Kushilevitz:97a}.  For example, with a single
bit of communication from $A$ to $B$, $\alpha(i,j)$ is independent of
$B$'s measurement $j$ and $\beta(i,j)$ can depend only on a partition
of the set of possible $i$'s into two sets.  Despite this
complication, deterministic protocols still correspond to a finite set
of vectors of probabilities $\vec{d}_\zeta^{\,(r)}$ in ${\mathbb
  R}^D$.

If we now allow randomness, the set of accessible probabilities
$\Omega_{MK}^{(r)}$ is given by the convex combination of the
deterministic probabilities, $\vec{p}= \sum_\zeta \lambda_\zeta
\vec{d}_\zeta^{\,(r)},~\sum_\zeta \lambda_\zeta =1,~\lambda_\zeta \geq
0$.  Again, $\Omega_{MK}^{(r)}$ is a convex combination of a finite
number of extreme points---a polytope---and can be described by a
finite set of facet inequalities: $ \vec{p} \in \Omega_{MK}^{(r)}~{\rm
  iff}~\vec{f}_\eta^{\,(r)} \cdot \vec{p} \leq c_\eta, \forall \eta$.
The complete set of facet inequalities for $\Omega_{MK}^{(r)}$ is a
complete set of Bell inequalities with $r$ bits of communication.  An
important limit arises when $r \geq 2 \log_2 M$ because then each
party can tell the other exactly which measurement they have selected.
In this setting, all deterministic protocols can be executed by the
two parties: the probability distribution $p_{a,b|i,j}$ is
unrestricted.  This implies that Bell inequalities with auxiliary
communication are trivial when $M = 1$.

Additionally, for $r \geq \log_2 M$, Bell inequalities with auxiliary
communication, although not necessarily trivial, are never violated by
probability distributions arising from local measurements on a shared
quantum state.  In fact this is true for any probability distribution
satisfying the {\em no--one-way--signaling
  conditions}~\cite{nosignal}, $p_{a|i,j} \equiv \sum_{b=0}^{K-1}
p_{a,b|i,j} = p_{a|i},$ is independent of $j$ for all $a$ and $i$: A's
marginal probability distribution is independent of B's choice of
measurement.  In such cases, it is sufficient that only A communicate
her measurement choice.  The simulation procedure is as follows: for
each of A's measurements $i$, the parties share a random variable
$\tilde a_i$ drawn from the probability distribution $\{a, p_{a|i}\}$
(i.e. $\tilde a_i = a$ with probability $p_{a|i}$).  Suppose A chooses
measurement $i$ and B chooses measurement $j$.  A outputs $\tilde a_i$
and sends her measurement choice $i$ to $B$.  B then outputs $\tilde
b_{\tilde a_i,i,j}$, where $\tilde b_{a,i,j}$ is drawn from the
probability distribution $\{b, p_{\tilde a_i,b|i,j}\}$.  (The roles of
A and B in the no--one-way--signaling conditions and protocol
may be reversed.)

{\em A complete set of Bell inequalities with auxiliary
  communication.}---Consider the simplest case $M=K=2$ and $r=1$ bit.
The polytope $\Omega_{2,2}^{(1)}$ is $12$ dimensional and has $112$
extreme vectors. Using both the primal-dual algorithm and the double
description method~\cite{facets} for facet enumeration, we find that
this polytope has $48$ facets. $16$ facets describe trivial
inequalities, $p_{a,b|i,j} \geq 0$ ($0 \leq i,j,a,b \leq 1$).  Another
$16$ facets are of the form
\begin{equation}
p_{a_1,b_1|0,0}+p_{a_2,b_2|0,1}+p_{a_3,b_3|1,0}+p_{a_4,b_4|1,1} \leq 2
\label{eq:fstineq}
\end{equation}
with $(a_1a_2a_3a_4) \in \{(0101),(1010),(0110),(1001) \}$ and $(b_1b_2b_3b_4)
\in \{ (0011),(1100),(0110),(1001) \}$.  The remaining $16$ facets are given by
\begin{equation}
p_{a,0|i,j} + p_{a,1|i,j} + p_{0,b|\bar{i},\bar{j}} + p_{1,b|\bar{i},\bar{j}} -
p_{a,b|i,\bar{j}} \geq 0 \label{eq:scdineq}
\end{equation}
($0 \leq i,j,a,b \leq 1$) where $\bar{0}=1$ and $\bar{1}=0$.  The
above inequalities completely describe the region of probability
distributions that can be created with one bit of communication.
There are probability distributions which violate these inequalities:
for example, if $p_{a,b|i,j}= \delta^a_j \delta^b_i$,
Eq.~(\ref{eq:fstineq}) with $(a_1a_2a_3a_4)=(0101)$ and
$(b_1b_2b_3b_4)=(0011)$ is maximally violated: substitution gives $4
\not\leq 2$.

It is straightforward to check that any probability distribution
satisfying the no-signaling conditions satisfies inequalities
Eq.~({\ref{eq:fstineq}) and Eq.~(\ref{eq:scdineq}).  Finally, consider
the probability distribution $p_{a,b|i,j}={1 \over 2} \left(
\delta^a_0 \delta^b_i+ \delta^a_j \delta^b_0 \right)$.
This probability distribution violates the no-signaling conditions
(in both directions), but satisfies Eq.~({\ref{eq:fstineq}) and
Eq.~(\ref{eq:scdineq}), thus indicating that these inequalities
are strictly stronger than no-signaling.

{\em A complete set of Bell inequalities with auxiliary communication
  for the joint observable.}---The above complete Bell inequalities
with auxiliary communication were used to bound the allowed
probabilities $p_{a,b|i,j}$ for protocols using a specified amount of
communication.  In quantum theory we are often interested not in all
of the probabilities for a measurement scenario, but only on the value
of a particular joint observable.  This simplifies our computational
task, because we may project the polytope $\Omega_{MK}^{(r)}$ onto a
lower-dimensional subspace and only enumerate the facets of the
projected polytope, as we shall explain in the following.  We term a
complete set of facet inequalities for this convex set a complete set
of Bell joint observable inequalities with auxiliary communication.
These generalize the CHSH inequality~\cite{Clauser:69a} to protocols
with communication.

Consider a measurement scenario with probabilities $p_{a,b|i,j}$ and
identify measurement outcomes with values of local observables.  The
joint observable for the $i$th and $j$th measurements of $A$ and $B$
is then defined as
\begin{equation}
c_{i,j}=\sum_{a=0}^{K-1} \sum_{b=0}^{K-1} A_a B_b p_{a,b|i,j},
\label{eq:obs}
\end{equation}
where $A_a$ and $B_b$ are the values of the local observable
corresponding to measurement outcomes $a$ and $b$, respectively.  As
for the full measurement scenario, we may list the components of the
joint observable to form a vector $\vec{c}$ in ${\mathbb R}^D$ with
$D=M^2$ (compare $D = M^2 (K^2 - 1)$ for the full probability
distribution).  For deterministic protocols with at most $r$ bits of
communication, the allowed functions $\alpha$ and $\beta$ are the same
as in the previous section, but now correspond to vectors with
components $c_{i,j} = A_{\alpha(i,j)} B_{\beta(i,j)}$.  Since the map
given by Eq.~(\ref{eq:obs}) is linear, the vectors corresponding to
joint observables accessible using randomness remain convex
combinations of the vectors accessible via deterministic protocols.

We now specialize to the scenario where each party has local $\pm
1$-valued observables ($K=2$) and they exchange $r = 1$ bit of
communication.  The joint correlation observable then has components
$c_{i,j}=p_{0,0|i,j}+p_{1,1|i,j}-p_{0,1|i,j}-p_{1,0|i,j}$.  If $M =
2$, we obtain only trivial inequalities $-1 \leq c_{i,j} \leq 1$.
That the inequalities are trivial also follow from
Eqs.~({\ref{eq:fstineq}) and (\ref{eq:scdineq}), for in this scenario
all possible joint observables can be obtained from probability
distributions that satisfy the no-signaling
conditions~\cite{poptsi}.

If $M=3$, the polytope has $320$ extreme vectors. Using both the
primal-dual algorithm and double description method for facet
enumeration~\cite{facets} we find that this polytope has $498$ facets.
$18$ of these describe the trivial inequalities $-1 \leq c_{i,j} \leq
1$.  The remaining $480$ facets can be described by the inequalities
\begin{equation}
\sum_{i,j=0}^2 M_{i,j} c_{i,j} \leq 1 \label{eq:corbel}
\end{equation}
where $M_{i,j}$ is either
\begin{equation}
M_1 = {1 \over 6} \left(
\begin{array}{ccc}
0 & -1 & 1 \\ -1 & 1 & 1 \\ 1 & 1 & 1 \end{array} \right),~M_2={1 \over 11} \left(
\begin{array}{ccc}
1 & 2 & -2 \\ 2 & 1 & 2 \\ -2 & 2 & 1 \end{array} \right),
\end{equation}
or any matrix obtained from these two matrices by (i) permuting the
rows and/or columns of the matrix and/or (ii) multiplying any subset
of the rows and columns of the matrix by $-1$.  The full set of
inequalities is a complete set of Bell joint observable inequalities
for $M=3$.

Let us show that quantum theory satisfies all of the above Bell joint
observable inequalities with auxiliary communication.  We do this for
a single one of the inequalities and the other inequalities all follow
by a similar argument.  For $\pm 1$ valued observables ${\bf A}_i$ and
${\bf B}_j$ and the joint quantum state $\bmath{\rho}$, a particular
inequality looks like ${\rm{Tr}} \left[ \bmath{\rho} {\bf T}_i \right]
\leq 1$, where ${\bf T}_i$ is the operator corresponding to matrix
$M_i$, e.g., ${\bf T}_1= [{\bf A}_1 (-{\bf B}_2 + {\bf B}_3) + {\bf
  A}_2 (-{\bf B}_1+ {\bf B}_2 + {\bf B}_3) +{\bf A}_3 ({\bf B}_1 +
{\bf B}_2 + {\bf B}_3 )]/6$.  ${\rm{Tr}} \left[ \bmath{\rho} {\bf T}
\right]$ is bounded by the sup norm of ${\bf T}$, $|{\bf
  T}|=\sup_{|\psi\rangle } \| {\bf T} |\psi\rangle \| / \| | \psi
\rangle \|$ and further ${\rm Tr}\left[ \bmath{\rho} {\bf T} \right]
\leq |{\bf T}^k | ^{1/k}$.  Calculation of ${\bf T}^k$ yields a
polynomial in ${\bf A}_i$, ${\bf B}_i$ and ${\bf I}$.  Since $|{\bf
  X}+{\bf Y}| \leq |{\bf X}| + |{\bf Y}|$ and $|{\bf P}| \leq 1$ for
any product ${\bf P}$ of ${\bf A}_i$, ${\bf B}_i$, and ${\bf I}$,
it follows that $|{\bf T}^k|$ is less than or equal to the sum of
the absolute value of the coefficients in the polynomial expansion
of ${\bf T}^k$.  By computer calculation we find that the sum of
the absolute value of the coefficients of ${\bf T}_1^4$ is
$\frac{155}{162}$ so that $|{\bf T}_1| \leq
\sqrt[4]{\frac{155}{162}}$.  Thus this Bell inequality with
auxiliary communication is satisfied.  Similar arguments using
${\bf T}_1^4$ or ${\bf T}_2^5$ show that all of the inequalities
Eq.~(\ref{eq:corbel}) are satisfied.  Therefore in the scenario
where each party chooses one of three two-outcome measurements, a
single bit of communication is sufficient to simulate the joint
correlation observable in quantum theory for all quantum states
and all quantum observables.

{\em Conclusion.}---Bell inequalities with auxiliary communication are
a powerful new tool for understanding the {\em cost} of producing
quantum correlations.  Surprisingly, in all the cases we considered,
it was sufficient to augment local realistic theories with a single
bit of communication to simulate the quantum correlations.  It remains
a challenge to find a Bell inequality with auxiliary communication
that is {\em violated} by a quantum state and set of quantum
measurements~\cite{note}.

 {\em Acknowledgements.}---We would like to acknowledge
conversations with Allison Coates and the comments of an anonymous
referee.  This work was supported by the National Science Foundation
under grant EIA-0086038.


\begin{thebibliography}{25}
\bibitem{note}
That such a violation must exist is implied by \cite{Brassard:99a}.
\end{thebibliography}
\end{document}